\documentclass[aps,prb,twocolumn,floats]{revtex4}
\usepackage{color,graphicx,pstricks}
\begin{document}  

\title {Electronic and Magnetic Reconstructions in Manganite Superlattices}

\author{Kalpataru Pradhan and Arno P. Kampf}

\affiliation{Center for Electronic Correlations and Magnetism, 
Theoretical Physics III, Institute of Physics, University of Augsburg, 
D-86135 Augsburg, Germany}

\date{\today}

\begin{abstract}
We investigate the electronic reconstruction at the interface between ferromagnetic metallic 
(FM) and antiferromagnetic insulating (AFI) manganites in superlattices using a two-orbital 
double-exchange model including superexchange interactions, Jahn-Teller lattice distortions, 
and long range Coulomb interactions. The magnetic and the transport properties critically 
depend on the thickness of the AFI layers. We focus on superlattices where the constituent 
parent manganites have the same electron density $n = 0.6$. The induced ferromagnetic moment 
in the AFI layers decreases monotonically with increasing layer width, and the 
electron-density profile and the magnetic structure in the center of the AFI layer gradually 
return to the bulk limit. The width of the AFI layers and the charge-transfer profile at the 
interfaces control the magnitude of the magnetoresistance and the metal-insulator transition 
of the FM/AFI superlattices.
\end{abstract}

\maketitle

\section{Introduction}

Correlated electron materials often involve the competition between various ordering tendencies 
of charge, orbital, spin, and lattice degrees of freedom. If further supplemented by weak 
disorder cluster coexistence, percolative transport, and colossal response phenomena emerge. 
It therefore remains a continuing challenge to understand the functional properties of 
transition-metal oxides (TMOs)\cite{dag-sci-tmo,tok-revs}. The physics at the surface of TMOs 
is further enriched by atomic and electronic reconstructions and complicated by the lack of 
inversion symmetry\cite{hwang-revs}. When surfaces of two TMOs were joined together to form an 
interface\cite{heber-sci}, new phenomena were discovered in the last decade. In some cases the 
phases at the interface are not even realized in the bulk of either of the TMOs which are 
joined together\cite{hwang-revs,triscone-revs-if}. For instance the discovery of the 
two-dimensional electron liquid\cite{ohtomo-electron-gas} which forms at the interface between 
the insulators LaAlO$_3$ and SrTiO$_3$ started a new subfield in the research on oxide interfaces. 
Subsequent experiments revealed superconductivity\cite{reyren-sc-if}, 
ferromagnetism\cite{brinkman-fm-if}, and even their coexistence\cite{dikin-fmnsc-if}. These 
unexpected phases at interfaces pose fundamental physics questions and simultaneously bear 
promises of technological importance for the design of novel materials\cite{manhart-revs-if}.

The perovskite manganese oxides are a particularly remarkable example for the mutual 
coupling of electronic and lattice degrees of freedom\cite{tok-book,chatterji-book,dagotto-book}. 
The manganites RE$_{1-x}$AE$_{x}$MnO$_{3}$, where RE and AE denote rare and alkaline 
earth elements, respectively, are known for their colossal magnetoresistance\cite{jin-cmr}.  
The various phases in manganites with charge, orbital, and magnetic order have been 
elaborated for different combination of RE and AE elements and doping regimes 
$x$\cite{tok-revs,tok-book,dag-njp,kajimoto-pd}. The recent development in superlattices 
has created yet another tool to explore electronic and magnetic phases of manganites.

La$_{1-x}$Sr$_{x}$MnO$_{3}$, a solid solution of LaMnO$_{3}$ (LMO) and 
SrMnO$_{3}$ (SMO) exhibits a number of phases depending upon the doping concentration $x$. 
In addition to A-type at ${x} = 0$ and G-type antiferromagnetism at ${x} = 1$\cite{x-0n1} 
also ferromagnetic and C-type antiferromagnetic (AF) phases exist at low 
temperature in an intermediate doping regime. For example La$_{0.67}$Sr$_{0.33}$MnO$_{3}$ 
is a ferromagnetic metal and La$_{0.33}$Sr$_{0.67}$MnO$_{3}$ is an AF 
insulator\cite{kajimoto-pd,x-pd-1,x-pd-2}. In manganite superlattices 2m layers of 
LMO are deposited on m layers of SMO\cite{lmo2-smo,lmo2-smo-apl} or vice versa\cite{lmo-smo2}. 
For $m = 1$, LMO$_{2m}$/SMO$_m$ and LMO$_m$/SMO$_{2m}$ are the superlattice counterpart of the 
La$_{0.67}$Sr$_{0.33}$MnO$_{3}$ and La$_{0.33}$Sr$_{0.67}$MnO$_{3}$ manganites, 
respectively. 

For LMO$_{2m}$/SMO$_m$ superlattices the uniform FM behavior of the solid solution is recovered 
for $m \leq 2$. A metal-insulator transition (MIT) occurs for $m \geq 3$ accompanied by a 
strongly modulated magnetic structure across the interface\cite{lmo2-smo,lmo2-smo-apl}. Similarly, 
LMO$_m$/SMO$_{2m}$ shows a C-type AF phase for $m = 1$ and $m = 2$ that exist in the solid 
solution counterpart La$_{0.33}$Sr$_{0.67}$MnO$_{3}$. The N\'{e}el temperature 250 K of the 
solid solutions increases to 320 K for $m = 1$. For $m \leq 2$, charge transfer from LMO to 
SMO may retain properties similar to the solid solutions. In these two scenarios, 
the charge transfer is likely confined to 1-3 unit cells. For $m > 2$, the increasing distance 
between the interfaces limits the charge transfer to the near vicinity of the interface. Due to 
this charge confinement LMO and SMO regain their respective parent phases away from the interface. 

Considerable progress has been achieved in describing the modulated magnetic 
structure\cite{millis-lmo-smo,satpathy-lmo-smo,brey-afi-afi,dag-mit-lmo2-smo,dag-lmo-smo-prb} 
and the MIT\cite{dag-mit-lmo2-smo} in the LMO/SMO superlattices. The standard two-orbital 
double exchange model for manganites was implemented to calculate the average electron density 
for each layer of the LMO/SMO superlattices; away from the interface LMO and SMO layers regain 
their initial electron densities. The electron-density profile across the interface is determined 
by electrostatics with an electron transfer from the LMO to the SMO side. The magnetic-profile 
for each layer follows the bulk phase diagram at the corresponding local density along with a 
weak proximity effect\cite{millis-lmo-smo,dag-mit-lmo2-smo}. 

In the initial studies of manganite superlattices\cite{cheng-tri-layer,
ink-lcmo-lcmo,li-lcmo-pcmo,jo-lcmo-lcmo} the primary intent was to improve the tunneling 
magnetoresistance\cite{fert-tmr,gruenberg-tmr} in a FM/AFI superlattice where the constituent 
parent FM and AFI manganites have the same electron density, $n = 1-x$. FM/AFI superlattices with 
different combinations of manganites (FM = LSMO, LCMO; AFI = PCMO, GdCMO; 
$n = 0.67$)\cite{li-lcmo-pcmo,mathur-lsmo-pcmo,lian-lcmo-gdcmo,das-lsmo-pcmo} 
have been studied in recent years. The observed induced ferromagnetic moment in the AFI layers 
depends on their width and the induced moment is tunable by an external magnetic 
field\cite{li-lcmo-pcmo,das-lsmo-pcmo,mathur-lsmo-pcmo1}. Remarkably the required external field 
is much smaller than the magnetic field required to induce ferromagnetism in the parent AFI 
compound. This results in a large magnetoresistance in the FM/AFI superlattices which are 
therefore candidate materials for colossal magnetoresistance at room temperature. Manganite 
trilayers can also be used to design an efficient spin valve\cite{brey-0.67-0.67}.
 
In this paper, we have studied in detail the electronic and magnetic reconstructions in FM/AFI 
superlattices at electron density $n = 0.6$ for different widths of the AFI layers. Electrons 
are transferred from the the FM to the AFI layers at the interface even though the initial electron 
density in the bulk FM and bulk AFI are equal. Our study confirms the underlying one-to-one 
correspondence between the density profile and the magnetic profile in the superlattices. With 
increasing AFI layer width, the induced ferromagnetic moment in the AFI layers decreases 
monotonically and drives an MIT in the superlattices. In addition to the large magnetoresistance 
our calculations explain the observed temperature crossover between positive and negative 
magnetoresistance and the two ferromagnetic transitions observed in the 
superlattices\cite{li-lcmo-pcmo,das-lsmo-pcmo}.

The paper is organized as follows: In Sec. II, we introduce the two-orbital model for bulk 
perovskite manganites. For superlattices we specify the essential modification by adding 
long-range Coulomb (LRC) interactions and briefly present the applied Monte Carlo technique. 
The parameter space of the FM/AFI manganite superlattices is discussed in Sec. III. Electronic 
and magnetic reconstructions at manganite interfaces are emphasized in Sec. IV while Sec. V 
is devoted to the MIT. Results with and without LRC interactions are compared in Sec. VI. 
In Sec. VII, various combinations of electron-phonon couplings for FM and AFI manganites are 
considered. The role of disorder at the interface is discussed in Sec. VIII, and conclusions 
are presented in Sec. IX. 

\section{Reference Model for Manganites}

We construct a two-dimensional model Hamiltonian for manganite superlattices composed of 
FM regions separated by AFI regions as 
\begin{eqnarray}
H = H_{FM} + H_{AFI} + H_{lrc}, 
\end{eqnarray}
where both $H_{FM}$ and $H_{AFI}$ have the same reference Hamiltonian
\cite{yunoki-hamil,dag-hamil-pr,sanjeev-x-0,kp-bsite-prl} 
\begin{eqnarray}
H_{ref} &=& \sum_{\langle ij \rangle \sigma}^{\alpha \beta}
t_{\alpha \beta}^{ij}
 c^{\dagger}_{i \alpha \sigma} c^{~}_{j \beta \sigma}
 - J_H\sum_i {\bf S}_i\cdot{\mbox {\boldmath $\sigma$}}_i
+ J\sum_{\langle ij \rangle} {\bf S}_i\cdot{\bf S}_j \cr 
&&- \lambda \sum_i {\bf Q}_i\cdot{\mbox {\boldmath $\tau$}}_i
+ {K \over 2} \sum_i {\bf Q}_i^2 - 
{\mu\sum_{i \alpha \sigma} c^{\dagger}_{i \alpha \sigma} c^{~}_{i \alpha \sigma}}. 
\end{eqnarray}
\noindent
Here, $c$ and $c^{\dagger}$ are annihilation and creation operators for
itinerant $e_g$ electrons and $\alpha$, $\beta $ refer to the two Mn-$e_g$ orbitals 
$d_{x^2-y^2}$ and $d_{3z^2-r^2}$ labelled as $a$ and $b$, respectively.
The kinetic energy part involves the nearest-neighbor hopping of $e_g$ electrons with 
amplitude $t^{ij}_{\alpha \beta}$ (~$t_{a a}^x= t_{a a}^y \equiv t$,~$t_{b b}^x= t_{b b}^y
\equiv t/3 $,~$t_{a b}^x= t_{b a}^x \equiv -t/\sqrt{3} $,~$t_{a b}^y= t_{b a}^y
\equiv t/\sqrt{3} $)\cite{dag-hamil-pr},~where $x$ and $y$ denote the spatial directions 
on a square lattice. $J_H$ is the Hund$'$s rule coupling between $t_{2g}$ spins ${\bf S}_i$ 
and the $e_g$ electron spin {\boldmath $\sigma$}$_i$, and $J$ is the AF superexchange 
between the $t_{2g}$ spins. $\lambda$ measures the strength of the electron-phonon 
coupling between the Jahn-Teller (JT) distortions, ${Q}_{2i}, {Q}_{3i}$ 
and the orbital pseudo spin operators  
${\tau}_{i2} = \sum_{\sigma} (c^{\dagger}_{i a \sigma} c^{~}_{i b \sigma} + 
c^{\dagger}_{i b \sigma} c^{~}_{i a \sigma}$), 
${\tau}_{i3} = \sum_{\sigma} (c^{\dagger}_{i a \sigma} c^{~}_{i a \sigma} -
c^{\dagger}_{i b \sigma} c^{~}_{i b \sigma}$). Here ${Q}_{2i}$ and ${Q}_{3i}$ are the 
distortions corresponding to the normal vibration modes of the oxygen octahedron that 
remove the degeneracy of the e$_g$ levels. The stiffness of the JT modes is denoted 
by $K$. The stiffness of the breathing mode 
distortion (${Q}_{1i}$), which couples to the local electron density is considerably larger 
than $K$ and therefore, in the adiabatic approximation the coupling to ${Q}_{1i}$ is 
neglected\cite{dag-hamil-pr} in the Hamiltonian Eq. 2. 

We treat all t$_{2g}$ spins and lattice degrees of freedom as 
classical\cite{class-ref1,class-ref2}, and measure energies in units of the Mn-Mn 
hopping $t_{aa}$ = $t$. In manganites $t$ is approximately 0.2-0.5 eV\cite{tok-book,satpathy-t}. 
The estimated value of the Hund$'$s coupling is 2 eV\cite{okimoto-hunds}, $i.e.$ much 
larger than $t$. For this reason we further adopt the limit 
$J_H \rightarrow \infty$, for which the $e_{g}$ electron spin perfectly aligns along 
the local $t_{2g}$ spin direction. The infinite Hund$'$s rule coupling then naturally
leads to redefine the spinless $e_{g}$ electron operator as 
  $c_{{i} \alpha} =
  \cos(\theta_{i}/2)c_{{i}\alpha \uparrow}
  + \sin(\theta_{i}/2)e^{-i\phi_{i}}c_{{i}\alpha\downarrow},$
where the polar angle $\theta_i$ and the azimuthal angle $\phi_i$ determine the 
t$_{2g}$ spin orientation. In terms of the redefined e$_g$ electron operators, 
the kinetic energy takes the simpler form
\begin{center}
$H_{\rm kin} = -\sum_{<ij>, \alpha \beta}
  {\tilde t}^{ij}_{\alpha \beta} 
  c_{{i} \alpha}^{\dag}c_{{j} \beta}$ ,
\end{center}
\noindent
where ${\tilde t}^{ij}_{\alpha \beta}$ is defined as
${\tilde t}^{ij}_{\alpha \beta}$ = $\Theta_{ij}t^{ij}_{\alpha \beta}$
with
  $\Theta_{ij} \!=\! \cos (\theta_{i}/2)\cos (\theta_{j}/2)
  \!+\! \sin (\theta_{i}/2)\sin (\theta_{j}/2)
  e^{-i(\phi_{i}-\phi_{j})}$.
The factor controls the magnitude of the hopping amplitudes due to the
different orientation of $t_{2g}$ spins at sites $i$ 
and $j$. 

We set $K=1$ without loss of generality, because if the lattice variable Q is 
replaced by $\sqrt{K}Q$, $\lambda$ can be simply renormalized as $\lambda/\sqrt{K}$. 
The length of the $t_{2g}$ spins is set to $|{\bf S}_i|=1$. In an external magnetic 
field a Zeeman coupling $H_{mag} = -{\bf h}\cdot\sum_i {\bf S}_i$ is added to the 
Hamiltonian. 

The reference `manganite model' $H_{ref}$ in 2D is constructed to reproduce
the correct sequence of magnetic phases in the bulk limit\cite{kp-bsite-prl,kp-bsite-epl}.
The different phases that are captured at low temperature  
are an orbitally ordered insulator at {\it x} = 0\cite{sanjeev-x-0}, a charge and orbital 
ordered ferromagnetic insulator at {\it x} = 0.25\cite{kp-bsite-epl}, a ferromagnetic metal 
window around {\it x} = 0.67, 
the CE charge and orbital ordered insulator around {\it x} = 0.50, a magnetic two 
dimensional A-type AF phase, and a G-type AF phase at {\it x} = 1.00. 

The average electron density of the constituent FM and AFI manganites in the FM/AFI 
superlattices is fixed by choosing the same chemical potential $\mu$ throughout the 
superlattice. The LRC interaction between all the charges, essential to control the amount of charge 
transferred across the interface is taken into account via a self consistent solution of the 
Coulomb potentials $\phi_i$ at the mean-field level by setting\cite{millis-mft-int}

\begin{eqnarray}
{\phi}_{i} = \alpha t \sum_{j\neq i} \frac{\langle n_{j} \rangle-Z_j}
{|{\bf R}_{i}-{\bf R}_{j}|}
\end{eqnarray}
in the long-range Coulomb part of the Hamiltonian,
\begin{eqnarray}
H_{lrc} &=& \sum_{i} \phi_i n_i.
\end{eqnarray}
\noindent
It is assumed that all the point charges $Z_j$ from the background ions are fixed and confined 
to the Mn sites. $\langle n_{j} \rangle$ refers to the $e_g$ electron density at the 
Mn site ${\bf R}_j$. The long-range interactions between the background point charges and the 
$e_g$ electrons are thereby combined with the charge-neutrality condition in the superlattice. 
Alternatively the Coulomb potentials $\phi_i$ can be determined self consistently by solving the 
Poisson equation\cite{yunoki-poisson-int} which is equivalent to solving the Eq. 3. 

The Coulomb interaction strength is controlled by the parameter $\alpha$ = $e^2$/$\epsilon$$at$ 
where $\epsilon$ and $a$ are the dielectric constant and the lattice parameter, respectively. 
This mean-field level set up is the minimal basis to study the charge transfer across the 
interface\cite{millis-lmo-smo,dag-lmo-smo-prb,brey-afi-afi}; it is also commonly used in the 
context of semiconductor interfaces\cite{luth-book}.

The background dielectric constant $\epsilon$ is order of 20 in 
manganites\cite{loidl-dielectric,cohn-dielectric}. It includes the lattice and the atomic 
polarizability contributions only. The dipolar contribution from the charge carrier motion and 
the associated screening is neglected to treat the absolute permittivities of FM and AFI manganites 
on equal footing\cite{cohn-dielectric}. For a lattice constant $a = 4 \AA$, $t = 0.5$ eV, 
and $\epsilon = 20$, the approximate value of $\alpha$ is 0.2. The screening length in two 
dimensions (2D) is larger than in three dimensions (3D)\cite{screening-2d-3d}. This is taken into 
account by choosing a larger dielectric constant in the 2D model ansatz. Specifically we use 
$\alpha=0.1$ in our calculations. The $\alpha$ values employed earlier in 3D model calculations 
were in the range $~$ 0.1 - 1\cite{millis-lmo-smo,dag-lmo-smo-prb,brey-afi-afi} while a much smaller 
value of $\alpha$ was used in 1D\cite{nagaosa-1d}. 

We apply an exact diagonalization scheme to the itinerant electron system for each 
configuration of the background classical variables of the t$_{2g}$ spins and the lattice 
distortions. The classical variables are annealed by starting from a random configuration. 
At each temperature the annealing requires at least 2000 system sweeps, by visiting every 
site of the lattice sequentially and updating the 
system by a metropolis algorithm. For each system sweep the computation time scales as 
O($N^4$) where N is the number of lattice sites. With thousands of annealing sweeps at 
various temperatures this procedure is time consuming.

We instead resort to a Monte Carlo sampling technique based on the `traveling cluster 
approximation' (TCA) \cite{tca-ref}. The TCA uses a moving cluster of size $N_c$ 
constructed around the site of the Monte Carlo update. The computational cost thereby decreases 
to O($NN^3_c$) and allows to treat system sizes up to $\sim 40\times 40$, with an $8\times 8$ 
moving cluster. The method and the associated transport calculations for
N = $24\times 24$ were extensively benchmarked and successfully applied 
in several earlier studies \cite{sanjeev-x-0,kp-bsite-prl,kp-bsite-epl}. At 
each system sweep we additionally solve for the Coulomb potentials ${\phi}_{i}$ 
in $H_{lrc}$ self consistently until the electron density at each site is converged. 
The maximum relative error for the convergence of the electron density is set to 0.01 
at low temperatures while it is relaxed up to 0.03 at higher temperatures. 
For each fixed set of parameters, the calculations start from ten different initial 
realizations of the classical variables. All the physical quantities are averaged over 
the results for these ten starting configurations. 

\begin{figure} [t]
\centerline{
\includegraphics[width=8.75cm,height=5.75cm,clip=true]{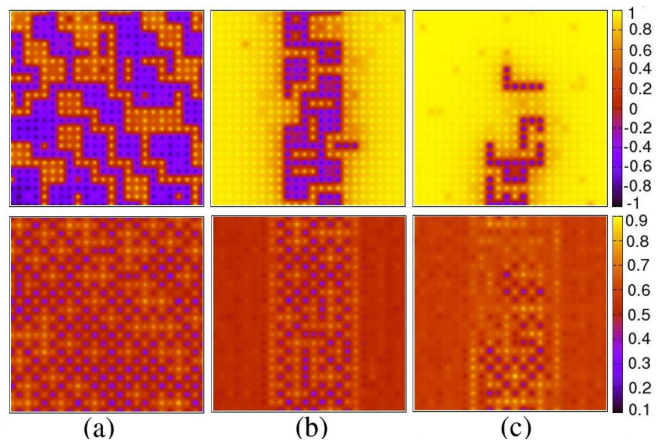}}
\caption{Color online:
1st row: The z components of the $t_{2g}$ spins; 2nd row: the electron-density for
each site on a 24$\times$24 lattice at T = 0.01. (a) The parent
AF insulator, (b) the superlattice with the width w = 11 of the AFI layer,
(c) the superlattice with w = 11 in the presence of an external magnetic
field, h = 0.004.
}
\end{figure}

\section{Superlattices of Ferromagnetic Metals and Antiferromagnetic Insulators}

Here we analyze specifically superlattices composed of FM and AFI manganites. 
Even with the above described 
simplifications in the model Hamiltonian we have to contend with manganite 
states at different doping $x$, AF superexchange strengths $J$, and different 
electron-phonon couplings $\lambda$. Also the proper choice of relevant parameters 
for the combination of ferromagnetic metals and AF insulators is not obvious. 
In addition, due to the slight structural and chemical mismatch between the parent FM and AFI 
manganites the parameters at the interface may be altered with respect to the 
bulk values. 

To constrain the parameter space we select the FM and AFI manganites of equal electron 
density $n$ = $1-x$. We choose $n$ = 0.6 above half filling to 
address the existing experiments on manganite superlattices near $x=1/3$. 
The AF superexchange couplings may vary depending on the different 
combinations of RE and AE ions. But for simplicity we will use the typical value 
$J = 0.1$\cite{perring-se} for both the FM and the AFI manganites. With the choice 
of the electron density and the superexchange couplings we are left with the 
crucial parameter $\lambda$ to differentiate between a FM and an AFI phase. 

Alternatively $\lambda$ can be fixed and the differentiation between the ferromagnetic 
metal and the AF insulator is achieved by varying $J$. This requires a larger value 
of $J$ in the AF insulator, but it leaves the ambiguity for 
choosing modified superexchange couplings at the interface. Yet another possibility is 
to keep each parameter $J$ and $\lambda$ fixed and to use different hopping 
amplitudes on the two sides of the interface. Again we are left with the possible 
modifications for the hopping amplitude at and across the interface. Since the 
$\lambda$ and $J$ values are measured in units of $t$, a smaller $t$ value 
is equivalent to the combinations of larger $\lambda$ and $J$ values or vice versa. 
Smaller and larger $\lambda$ values are thereby closely related to the larger and 
smaller bandwidth manganites, respectively.

\begin{figure} [t]
\centerline{
\includegraphics[width=8.75cm,height=3.75cm,clip=true]{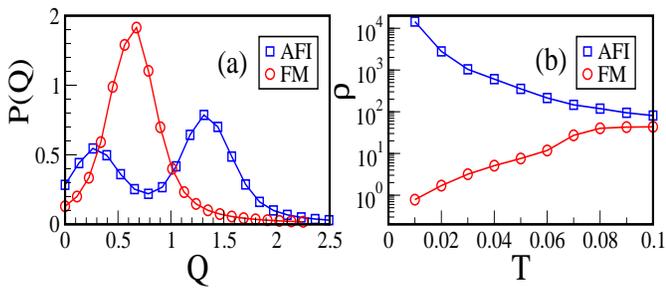}}
\caption{Color online: Bulk system at $n$ = 0.6 (a) Distribution function for the 
lattice distortions P(Q) of the bulk AFI ($\lambda_I=1.75$) and bulk 
FM ($\lambda_M=1.50$) at T = 0.01. (b) The resistivity variation with 
temperature for AFI and FM bulk systems.}
\end{figure}

Here we consider two $\lambda$ values to differentiate between the FM and the AFI 
manganites. For the parameters $J = 0.1$ and the density $n = 0.6$, the groundstate 
is FM for $\lambda \equiv \lambda_M = 1.50$ while it is an AFI for 
$\lambda \equiv \lambda_I = 1.75$. The AFI phase at $n = 0.6$ is not a perfect 
charge and orbital ordered CE phase which is stable only at 
$n = 0.5$\cite{dag-hamil-pr,kp-bsite-prl}. The AFI phase at $n = 0.6$ is inhomogeneous, 
it contains the CE-type zigzag ferromagnetic chains and local charge ordered regions 
as shown in Fig.1(a). 

We compute the `one point' distribution function of the lattice distortions, 
$P(Q)=\langle {1 \over N} \sum_i\delta(Q-\vert Q_i \vert) \rangle$ where N is the 
total number of sites, to compare them with the electron-density variations. 
The two peaks in Fig.2(a) for the AFI at T = 0.01 indicate that the distortions 
have a bimodal distribution. The sites with a large distortion attract more electrons 
with a density 
increase to n$_i$ $\sim$ 0.8. So the spatial density pattern in Fig.1(a) reflects 
simultaneously the large and small structural distortions. The single peak structure 
of P(Q) for the FM indicates instead that there are distortions of similar strength 
on all sites, which facilitates the motion of the electrons and thereby supports 
the metallic character.

\begin{figure} [t]
\centerline{
\includegraphics[width=8.75cm,height=3.75cm,clip=true]{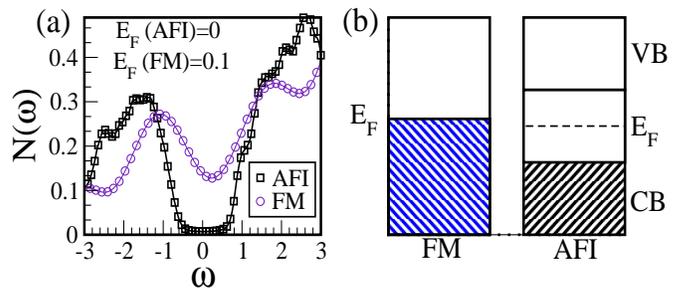}}
\caption{Color online: Bulk system at $n = 0.6$ (a) DOS of the bulk AFI 
($\lambda_I=1.75$) and FM ($\lambda_M=1.50$) at T = 0.01. (b) Schematic 
energy diagram.}
\end{figure}

We obtain the resistivity for both FM and AFI bulk phases by calculating the 
d.c. limit of the longitudinal conductivity as determined by the Kubo-Greenwood 
formula\cite{mahan-book}
\begin{equation}
\sigma ( \omega)
= {A \over N}
\sum_{\alpha, \beta} (n_{\alpha} - n_{\beta})
{ {\vert f_{\alpha \beta} \vert^2} \over {\epsilon_{\beta}
- \epsilon_{\alpha}}}
\delta(\omega - (\epsilon_{\beta} - \epsilon_{\alpha})),
\end{equation}
with $A = {\pi  e^2 }/{{\hbar a}}$ and $n_{\alpha}= \theta(\mu - \epsilon_{\alpha})$, 
$a$ is the lattice spacing. $f_{\alpha \beta}$ denotes the matrix elements of the 
paramagnetic current operator $j_x = i t  \sum_{i, \sigma} (c^{\dagger}_{{i + x},\sigma}
c_{i, \sigma} - h.c.)$ between eigenstates $\vert \psi_{\alpha}\rangle$,
$\vert \psi_{\beta}\rangle$,  and $\epsilon_{\alpha}$, $\epsilon_{\beta}$
are the corresponding eigenenergies. We extract the d.c. conductivity by calculating 
the `average' conductivity for a small low frequency interval $\Delta \omega$ defined as
\begin{equation}
\sigma_{av}(\Delta \omega)
= {1 \over {\Delta \omega}}\int_0^{\Delta \omega}
\sigma(\omega){\rm d} \omega . \nonumber
\end{equation}
\noindent  
$\Delta \omega$ is chosen two to three times larger than the 
mean finite-size gap of the system as determined by the ratio of the bandwidth and 
the total number of eigenvalues. This procedure has been benchmarked 
in a previous work\cite{cond-ref}. Fig.2(b) shows the temperature dependence of the 
resistivity $\rho = \sigma_{av}^{-1}$ in units of ${\hbar a}/{\pi  e^2 }$. 
The insulating behavior in the AFI at low temperatures is due to charge and orbital 
order, which opens an energy gap in the spectrum. 

The density of states (DOS) 
$N(\omega)=\langle {1 \over N} \sum_\alpha \delta(\omega-\epsilon_\alpha) \rangle $ is 
shown in Fig.3(a). The center of the gap in the DOS of the AFI state is chosen as the 
energy zero. With this choice the Fermi energy of the FM state at the same density is 
at $\epsilon_F=0.1$. The DOS is finite for the FM state at its Fermi level. Fig.3(b) 
translates the results for the DOS into a schematic energy diagram. When the FM and 
the AFI manganites are joined together band bending near the interface is expected to 
shift electrons from the FM to the AFI side. An FM/AFI superlattice is shown 
schematically in Fig.4. The width of the middle AFI layer, sandwiched between the FM 
layers is denoted by w. Periodic boundary conditions are enforced in both directions 
and thereby represent a superlattice structure composed of alternating FM and 
AFI regions. 

\begin{figure} [t]
\centerline{
\includegraphics[width=5.75cm,height=5.75cm,clip=true]{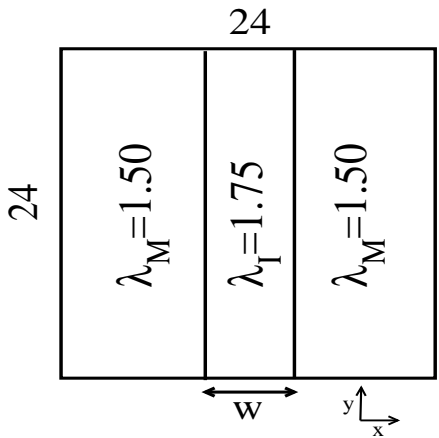}}
\caption{Schematic view of the FM/AFI superlattice on a 24$\times$24 lattice.
We consider periodic boundary conditions in both directions.}
\end{figure}

\section{Charge Transfer and magnetic reconstruction at the Interface}

As we show below, if the width of the AF regions is small, they may become ferromagnetic 
and metallic in the superlattice structure. The induced magnetization is restricted 
to the near vicinity of the interfaces. The distance from the interface at which the 
parent AF insulating character is recovered depends upon the parameter $\alpha$ and 
the electron-phonon couplings $\lambda_M$ and $\lambda_I$ . 

In this section we start with the specific choice of $\lambda_M=1.50$, $\lambda_I=1.75$, 
and $\alpha=0.1$, and discuss results for the decrease of the magnetization with 
increasing width 
of the AFI layers. This decrease is quantified by the average z component of the 
t$_{2g}$ spins $\langle S_z \rangle_I$ in the AFI layer, where the angular bracket 
denotes the thermal average combined with an additional average over ten different 
`samples'. We also define a measure for the local staggered charge order by 
$\langle CO \rangle _I$ = ${1 \over N_I}$ $\sum_{i \in AFI}$ $\langle n_i \rangle$
e$^{i(\pi,\pi) \cdot {{\bf r}_i}}$ for the AFI layer; $i$ denotes one of the $N_{I}$ 
lattice sites in the AFI layer with position ${{\bf r}_i}$. In Fig.5(a), we plot 
$\langle S_z \rangle _I$ and $\langle CO \rangle _I$ for different widths w. For 
w $\leq$ 5, the averaged 
$\langle S_z \rangle_I$ is near unity and starts to decrease for w $>$ 5, while 
$\langle CO \rangle _I$ remains small for w $\leq$ 5 and starts to rise for w $>$ 5. 
Similarly (see Fig.5(b)) also the long-wavelength magnetic structure factor 
$S_I(\textbf{0})$ starts to decrease beyond w = 5, where 
$S_I(\textbf{q})$ =${1 \over N_I^2}$ $\sum_{ij \in AFI}$ $\langle \bf {\bf S}_i 
\cdot {\bf S}_j \rangle$ e$^{i\textbf{q} \cdot ({\bf r}_i-{\bf r}_j)}$. 
So the decrease in the magnetization beyond a critical width w$_c$ is accompanied 
by emerging charge order in the AFI layer. The critical width for fixed $\lambda_I$ 
may change for a different set of parameters $\lambda_M$ and $\alpha$. This will be 
discussed in sections VI and VII. The critical width is therefore not directly related to 
an intrinsic length scale of the parent bulk AFI. 

The averaged magnetization necessarily does not reveal any spatial variation 
transverse to the interface. In order to analyze the magnetization profile, 
we calculate the $\langle S_z(x) \rangle$ for each line of the superlattice 
with transverse coordinate $x$. Fig.6(a) shows the line averaged 
$\langle S_z(x) \rangle$ vs. line index for w = 7. $\langle S_z(x) \rangle$ 
decreases for $x$ = 12, 13, and 14, i.e. in the center lines of the AFI layer. 
So the induced ferromagnetic moment in the AFI layer is indeed confined to 
just the two lines nearest to the interface. 

In the bulk for $\lambda_I=1.75$ the AFI phase extends to electron densities larger 
than $n = 0.6$ and changes to a ferromagnetic insulating phase at $n=0.75$\cite{kp-bsite-epl}. 
It is therefore natural to explore the connection between the induced magnetization in 
the AFI layer with the electron densities at the interface. As shown in Fig. 6(b), at 
the interface the line averaged electron density $\langle n(x) \rangle$ of the FM layer 
decreases while $\langle n(x) \rangle$ for the AFI layer increases beyond the initial 
electron density 0.6. In fact $\langle n(x) \rangle$ at the fully magnetized lines $x=10$ 
and 16 is nearly equal to 0.75. 

\begin{figure} [t]
\centerline{
\includegraphics[width=8.75cm,height=3.75cm,clip=true]{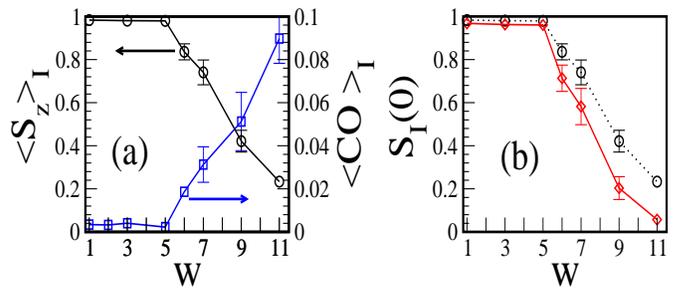}}
\caption{Color online: The average z component of t$_{2g}$ spins, 
$\langle S_z \rangle_I$ 
and the average staggered charge order $\langle CO \rangle _I$ (see text) in the 
AFI layer at T = 0.01.
(b) Ferromagnetic structure factor  S$_I$({\bf 0}) for different widths of 
the AFI layer at T = 0.01. The average z component is also re-plotted as the dotted 
line.
}
\end{figure}

The direction of electron transfer from the FM to the AFI regions was anticipated 
already in the discussion in Section III. Away from the interface the average electron 
density gradually returns to the initial electron density $n = 0.6$. In addition to the 
overall charge transfer, $\langle n(x) \rangle$ is spatially modulated perpendicular 
to the interface. These charge modulations are Friedel-like density oscillations; 
their amplitude decreases with the distance from the interface. The spatial density 
variations are nearly symmetric around the central line, $x=13$. As we have verified 
a perfect symmetry can be achieved by averaging over a larger number of samples.  

For the incommensurate filling $n = 0.6$, the parent AFI is fragmented into small regions 
with and without charge and orbital order. High-density sites (see Fig.1(a)) are always 
accompanied by large lattice distortions. Similarly a one-to-one correspondence between 
the the electron densities and the lattice distortions exists in the superlattice. The 
line averaged lattice distortions $\langle Q(x)\rangle$ (see Fig.6(c)) are infact larger 
in the AFI layer. $\langle Q (x)\rangle$ is modulated similarly to the line averaged 
electron density shown in Fig.6(b). 

\begin{figure} [t]
\centerline{
\includegraphics[width=8.75cm,height=7.75cm,clip=true]{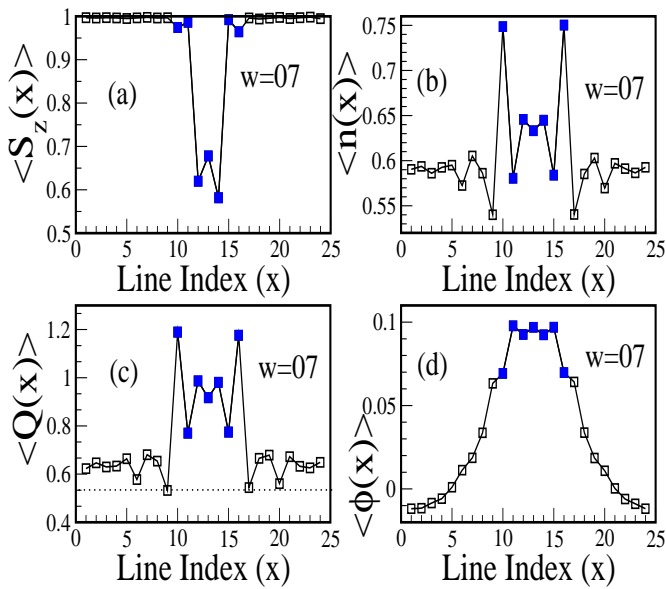}}
\caption{Color online: Line averaged (a) z component of the t$_{2g}$ spins 
$\langle S_z(x) \rangle$, (b) electron density $\langle n(x) \rangle $, 
(c) LRC potentials $\langle \phi(x) \rangle$, and (d) lattice distortions 
$\langle Q(x) \rangle$ vs. line index for w = 7. Open and closed symbols
are for the FM and the AFI layers, respectively. The temperature is 
T = 0.01}
\end{figure}

The amount of charge transfer from the FM to the AFI layers is controlled by 
the LRC interactions. The line averaged Hartree potential $\langle \phi (x)\rangle$ 
in Fig.6(d) is negative for $x \leq 4$ and $x \geq 22$. The average electron densities 
for those lines is nearly equal to the initial electron density $n = 0.6$. The charge 
transfer is restricted to a few lines near the interface where the averaged potentials 
$\langle \phi(x) \rangle$ are positive and thereby counteracts further electron transfer. 
Since $\phi_i$ and $n_i$ are the self-consistent solutions of the Poisson equation, the 
increase or the decrease of the averaged electronic density has a one-to-one 
correspondence to the negative or positive curvature of $\langle \phi(x) \rangle$.
 
For $\alpha=0.1$, see Fig.7(a), the electron densities are clearly lower (higher) 
in the AFI (FM) layer as compared to $\alpha=0$. The LRC potentials decrease the 
electron density in the center of the AFI layer, but the average electron densities 
for individual lines at $x=12$, $x=13$, and $x=14$ are still enhanced due to the 
Friedel-like density oscillations. The electron density 
in the central line $x=13$ results from the superposition of the oscillations 
originating from $x=10$ and $x=16$. The constructive superposition is more 
pronounced for the AFI layer width w = 5 shown in Fig.7(b), for which the center 
line is fully magnetized due to the enhanced electron density and the induced 
magnetism originating from the neighboring magnetized 2+2 interfacial lines. For w = 3, 
in Fig.7(c), the electron density of the central line is nearly equal to the 
initial electron density and the charge transfer is truly confined to the 
interface. For w $>$ 7, density oscillations in the AFI layer disappear due the 
recovery of the bulk AFI character (see Fig.7(d) for w = 9). 

\begin{figure}
\centerline{
\includegraphics[width=8.75cm,height=7.75cm,clip=true]{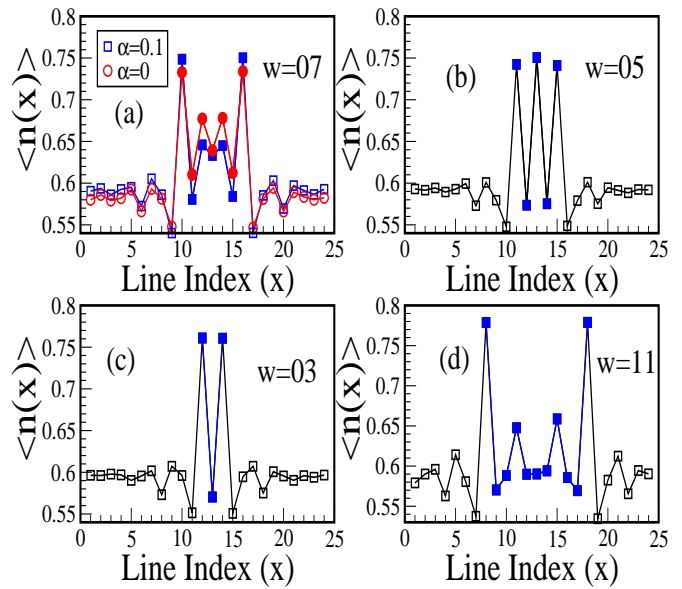}}
\caption{Color online: Line averaged electron density $\langle n(x) \rangle$
for (a) w = 7 with ($\alpha=0.1$) and without ($\alpha=0$) LRC interaction, 
(b) w = 5, (c) w = 3, (d) w = 11. Open and closed symbols are for the FM and the 
AFI layers, respectively. The temperature is T = 0.01}
\end{figure}

\section{Metal-Insulator transition}

In this section we compare the magnetic structure factor $S_I({\bf 0})$ in the AFI layer   
with the resistivity for the superlattice for current flow perpendicular to the interface. 
Fig.8(a) shows the temperature dependence of $S_I({\bf 0})$ in the AFI layer for 
different layer width w. For all temperatures, the ferromagnetic structure factor 
and the onset temperature for ferromagnetism decrease with increasing w. The z component 
of the t$_{2g}$ spins $\langle S_{zi} \rangle$ and the electron densities for each
site in the superlattice are shown in Fig.1 (b) for w = 11. The two lines in the AFI layer 
nearest to the interface are aligned ferromagnetically to the t$_{2g}$ spins on the 
FM side in accordance with the discussion in Section IV. The magnetic structure of the 
center lines is already similar to the bulk AFI phase, and the density profile reveals 
the emergence of local charge ordering patterns in the AFI layer. The superlattice for 
w = 11 is therefore likely to be an insulator.

\begin{figure}
\centerline{
\includegraphics[width=8.75cm,height=7.75cm,clip=true]{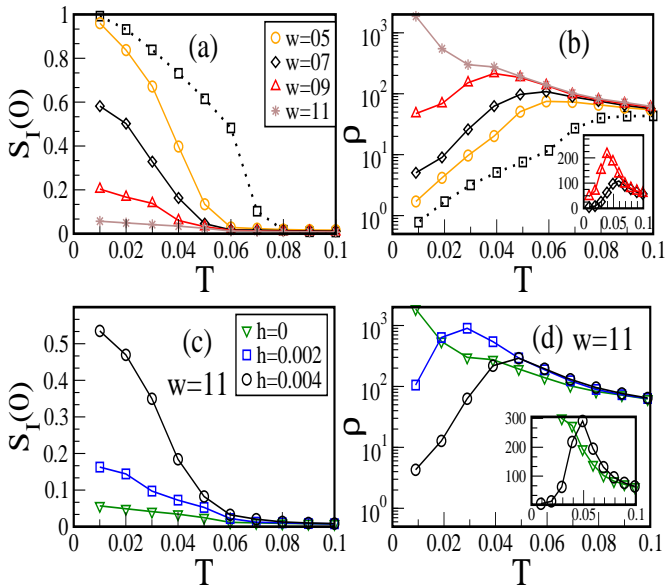}}
\caption{Color online: Temperature dependence of (a) the ferromagnetic
structure factor $S_I({\bf 0})$ and (b) the resistivity $\rho$ for different widths 
of the AFI layer. The dotted lines in (a) and (b) are for the bulk FM. 
(c) $S_I({\bf 0})$  and (d) resistivity variation with temperature for 
w = 11 in the presence of the external magnetic fields h = 0.002 and 0.004. 
Insets in (b) and (d) show the resistivity variation in the linear temperature 
scale.
}
\end{figure}

Fig.8(b) displays the temperature dependence of the longitudinal resistivity $\rho$ of 
the superlattice. $\rho$ for w = 11 steeply rises towards low temperatures as expected 
for an insulator. For w $\leq$ 7, the superlattice is ferromagnetic and metallic at low 
temperatures. For w = 5, the onset temperature for ferromagnetism in the AFI layer is 
near 0.05, and simultaneously the resistivity start to decrease. 

The low-temperature resistivity increases monotonously with the AFI layer width. An MIT 
occurs beyond a threshold width similar as in the experiments by 
Li {\it et al.}\cite{li-lcmo-pcmo} who measured the variations in the resistivity in 
LCMO/PCMO superlattices for different widths of the PCMO layer. In addition, the inset in 
Fig.8(b) shows a hump in the resistivity for w = 9 and w = 7 at the ferromagnetic onset 
temperatures T$\simeq$ 0.04 and 0.05, respectively, while there is no such hump in the bulk 
FM manganite (see the dashed line of Fig.8(b)). The humps in the resistivity curves increase 
and shift to lower temperatures with increasing AFI layer width (for w $\le$ 9). This trend 
agrees well with the LCMO/PCMO superlattice experiment\cite{li-lcmo-pcmo} with the exception 
of the experimentally observed hump in the bulk FM limit. The temperature 
dependence of $\rho$ near the MIT in bulk FM manganites is tied to the presence of intrinsic 
inhomogeneities and disorder\cite{burgy-mit,motome-mit}. Here we have not included any 
disorder\cite{kp-bsite-prl} in the Hamiltonian; the MIT in the FM/AFI superlattice has a 
different origin. It is due to the induced onset of ferromagnetism in the AFI layer. 

The natural next step is to explore the effect of an external magnetic field on the MIT in 
the superlattice. Fig.8(d) shows that the resistivity for w = 11 is lowered by the magnetic 
field at low temperatures. The connection between the resistivity decrease and the onset of 
ferromagnetism in the AFI layer is verified in Fig.8(c). The MIT for h = 0.004 is at 
T $\simeq$ 0.05; above this temperature, the resistivity is larger than the resistivity 
calculated for h = 0. The magnetoresistance, $[\rho(h)-\rho(0)]/\rho(0)$ therefore changes sign 
near the onset temperature for ferromagnetism. 

At higher temperatures, the t$_{2g}$ spins are randomly oriented. We recall that in the 
limit $J_H \rightarrow \infty$ the $e_{g}$ electron spins are perfectly aligned along the 
local $t_{2g}$ spin direction. So the current is equally carried by up-spin and down-spin 
electrons. In the presence of a weak external magnetic field, the number of $e_{g}$ electrons 
is increased in the up-spin channel only in the FM layers and the tunneling current from the 
down-spin channel is decreased. The up-spin channel in the AFI layer remains unaltered and 
restricts the possible enhancement of the tunneling current from the up-spin channel. This 
results in an overall increase of the resistivity and a positive magnetoresistance. 

At low temperatures in an external magnetic field the t$_{2g}$ spins in the AFI layer tend to 
align in the same direction as the t$_{2g}$ spins in the FM layers. So the tunneling current 
in the down-spin channel decreases but it increases for the up-spin channel. This decreases 
the overall resistivity of the superlattice. The corresponding crossover of the magnetoresistance 
from negative to positive was indeed reported in 
La$_{0.7}$Ce$_{0.3}$MnO$_3$/SrTiO$_3$-Nb\cite{sheng-mr-pn} and 
La$_{0.82}$Ca$_{0.18}$MnO$_3$/SrTi$_{0.95}$Nb$_{0.05}$O$_3$\cite{zhou-mr-pn} heterojunctions. 
  
The z component of the t$_{2g}$ spins and the electron densities for each site in the 
superlattice are shown in Fig.1(c) for w = 11 and h = 0.004. Ferromagnetic and AF structures 
coexist spatially separated in the AFI layer. The ferromagnetic phase in the upper panel 
is directly connected to the charge disordered regions in the lower panel. A magnetic field 
h $\geq$ 0.007 is required to tune the parent AFI into a ferromagnetic metal. Lower fields are 
instead required for the FM/AFI superlattices which are therefore candidates to design materials 
with large magnetoresistance. 

\section{Results without LRC interactions}

In order to analyze the effect of the LRC interactions we examine the ferromagnetic structure 
factor S$_I$({\bf 0}) for $\alpha=0$ in Fig.9(a). The critical width, beyond which S$_I$({\bf 0}) 
starts to decrease, is w$_c$ = 7. With the LRC potentials, $\alpha=0.1$, the critical width 
w$_c$ = 5, as discussed in section IV, is smaller due to the decrease in the average electron 
density in the AFI layer which is evident in Fig.7(a). All the discussions in the previous two 
sections remain qualitatively valid also for $\alpha=0$, but with a larger critical width w$_c$.

\begin{figure}
\centerline{
\includegraphics[width=8.75cm,height=7.75cm,clip=true]{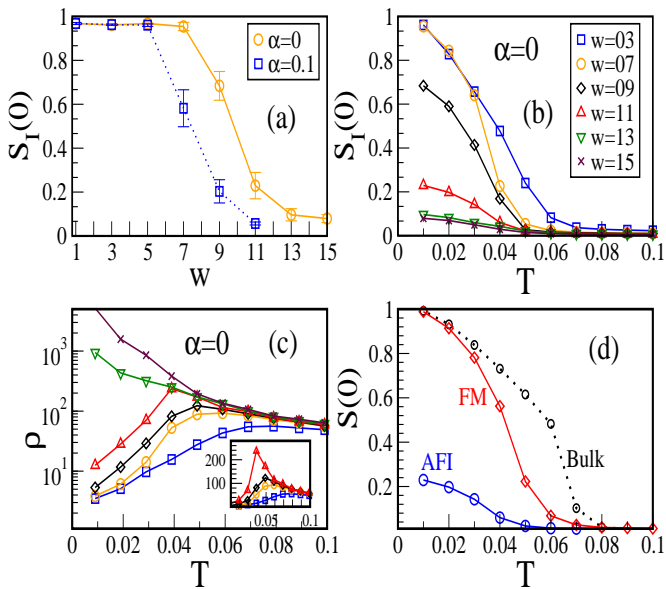}}
\caption{Color online: For $\alpha=0$ (a) $S_I({\bf 0})$ for different 
widths of the AFI layers at T = 0.01. $S_I({\bf 0})$ for $\alpha=0.1$ is also 
re-plotted as the dotted line. Temperature dependence of 
(b) $S_I({\bf 0})$ and (c) the resistivity for different 
widths of the AFI layers. Legends in (b) and (c) are the same. Inset in (c) shows 
the temperature dependence of the resistivity in the linear scale. 
(d) The temperature dependence of ferromagnetic structure factor for the 
AFI layer, the FM layer for w = 11, and the bulk FM. 
}
\end{figure}

In Fig.9(b), we plot the temperature dependence of $S_I({\bf 0})$ in the AFI layer for $\alpha=0$. 
Similar to the results in Fig.8(a), the onset temperature for ferromagnetism decreases with 
increasing w. Fig.9(c) shows the temperature dependence of the longitudinal resistivity of the 
superlattice. For w = 13 and w = 15, the low-temperature resistivity rises by orders of magnitude. 
The inset of Fig.9(c) zooms into the MIT at intermediate temperatures on a linear scale. The humps 
in $\rho$(T) decrease and shift to higher temperatures for w $<$ 11. 

The onset temperature of ferromagnetism is higher in the FM layers than the AFI layers as becomes 
obvious from the temperature dependent structure factor in Fig.9(d). So there are two ferromagnetic 
transitions in the superlattice: the first at T$_{c1}$ for ferromagnetism in the FM layers and the 
second for the global ferromagnetism at T$_{c2}$ $<$ T$_{c1}$. The rise in the resistivity $\rho$ near 
the hump just below T$_{c1}$ is apparently due to the onset of ferromagnetism in the FM layers. 
The rise in $\rho$ is expected for the same reason as for positive magnetoresistance discussed in 
section V. The downturn in $\rho$ results from the onset of global ferromagnetism in the 
superlattices at T$_{c2}$. These two transition temperatures are also observed in the LCMO/PCMO 
superlattices\cite{li-lcmo-pcmo}. The temperature dependence of the ferromagnetic structure factor 
in the bulk FM state is included as the dotted line in Fig.9(d). The ferromagnetic onset temperature 
for the FM layers in the FM/AFI superlattices is lower than for the bulk.

\section{Variation of \boldmath${\lambda_M}$ and \boldmath${\lambda_I}$}

In this section we compare $S_I({\bf 0})$ for different combinations of electron-phonon couplings 
$\lambda_M$ and $\lambda_I$. In the previous sections $\lambda_M$ = 1.5 and $\lambda_I$ = 1.75 
were chosen. If $\lambda_I$ is kept fixed and $\lambda_M$ varies between 1.0 and 1.6, 
the groundstate is ferromagnetic and metallic in the bulk limit at $n = 0.6$.
The induced magnetization in the AFI layer changes little for $\lambda_M \leq 1.5$ 
and for different widths w (see Fig.10(a)). For $\lambda_M$ = 1.6 however, $S_I({\bf 0})$ 
is reduced for w $\geq$ 5. The difference results from the 
decrease in the electron density in the FM layers nearest to the
interface as becomes evident from the comparison with the bulk phase at 
$n=0.5$\cite{kp-bsite-prl}. For $\lambda_M \leq 1.5$, the groundstate at
$n=0.5$ is a ferromagnetic metal while it is a CE-type antiferromagnet for 
$\lambda_M = 1.6$. For $\lambda_M \leq 1.5$, the magnetization in the interfacial 
FM layers is not
altered, even if the electron density decreases towards $n=0.5$. But 
for $\lambda_M=1.60$, ferromagnetic order becomes unstable at the interfacial lines. 
The smaller magnetization at the interfacial FM layers 
decreases the induced magnetization in the AFI layer for w $\geq$ 5. 

\begin{figure}
\centerline{
\includegraphics[width=8.75cm,height=4.00cm,clip=true]{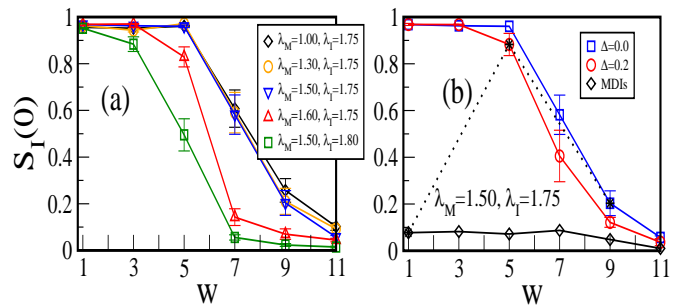}} 
\caption{Color online: $S_I({\bf 0})$ for different combinations of $\lambda_M$ and 
$\lambda_I$ values at T = 0.01. (b) $S_I({\bf 0})$ without ($\Delta=0$) and with 
($\Delta=0.2$) quenched disorder, and with magnetically disordered interfaces (MDIs) 
(see text) for $\lambda_M=1.50$ and $\lambda_I=1.75$ at T = 0.01. The dotted line 
join the three points namely w = 1 (for MDIs), w = 5 (for $\Delta=0.2$), and w = 9 
(for $\Delta=0$).}
\end{figure}

Fig.10(a) also indicates the reduced magnetization in the AFI layer for the larger 
electron-phonon coupling $\lambda_I=1.8$, for which the AFI layer recovers the AF, charge 
ordered state at a smaller width w. It is therefore easier to induce a ferromagnetic moment 
in large-bandwidth (small $\lambda$) manganites. At $n = 0.67$, the bandwidth of PCMO is the 
largest among those manganites for which an AF, charge ordered insulating phase is 
experimentally observed\cite{kajimoto-pd}. In fact, in most of the experimental FM/AFI 
superlattices at $n = 0.67$ the insulating manganite PCMO is used along with a variety of 
different FM manganites. 
 
\section{Disorder at the Interface}

The size mismatch of RE and AE elements in the manganites RE$_{1-x}$AE$_x$MnO$_3$ leads 
to tilting and distortions of the MnO$_6$ octahedra and variations in the local electronic 
parameters. The tilting and distortions of MnO$_6$ octahedra is generally known as A-type 
disorder. Here we have neglected the intrinsic A-type disorder in both the FM and the AFI 
manganites. Their interface is more prone to disorder due to chemical intermixing, lattice 
mismatch, and A-type disorder. We test the effect of quenched binary disorder in the 
terminating line of the FM layers at the interface by adding the potential disorder term 
by $\sum_j \epsilon_j n_j$ to the Hamiltonian. The sum over j is restricted to the 
terminating lines of the FM layers and $\epsilon_j$ is the quenched disorder potential with 
$\overline{\epsilon}_j=0$ and values $\pm$$\Delta$. 

In Fig.10(b), we show the induced magnetization in the AFI layer for $\Delta=0.2$ and 
different widths w, which apparently changes little between $\Delta=0$ and $\Delta=0.2$. 
All the discussions in the previous sections remain qualitatively unchanged for moderate 
disorder $\Delta=0.2$. The magnetization of the interfacial lines of the FM layers, where 
$\Delta$ is included, is nevertheless likely to decrease for larger $\Delta$. Ultimately 
the magnetization in those interfacial lines will be quenched due to spin disorder. In order 
to see the effect of the magnetically disordered interfaces on the AFI layer we have 
fixed randomly oriented spins in the interfacial lines throughout the Monte Carlo simulations. 
Indeed, in this extreme case, the induced magnetization in the AFI layer is very small 
(see Fig.10(b)), irrespective of the AFI layer width. The magnetization of the FM layer at 
the interface is therefore crucial to induce the ferromagnetic moments in the AFI layer. 

In the experiments it is apparently not clear whether or not there are magnetically 
disordered interfaces on the FM sides in the FM/AFI superlattices\cite{mathur-lsmo-pcmo}. 
Thin AFI layers are likely to be more susceptible to disorder due to strain effects. 
The disorder strength $\Delta$ at the interface will therefore increase with decreasing 
the AFI layer width; the width dependence of the disorder strength is however hard to 
quantify. If we combine the results in Fig.10(b) such that disorder strength is maximum 
for smaller AFI layers width and decreases thereafter for larger w (see the dotted line) 
then the magnetic moment in the AFI layer behaves non-monotonically which agrees 
qualitatively with the LSMO/PCMO superlattice experiment\cite{mathur-lsmo-pcmo}. But 
more theoretical work is needed to understand this non-monotonic behaviour. 

\section{Conclusions}

We have investigated the magnetic and electronic properties of manganite superlattices 
at the specific electron density $n = 0.6$ using a two-orbital double-exchange 
model including super-exchange interactions, JT distortions, and LRC interactions 
in 2D. At the interface electrons are transferred from the the FM to the 
AFI manganites. Friedel-like density oscillations are observed in both the FM and the 
AFI layers. Due to the charge transfer and the induced magnetization the smaller width 
AF insulator sandwiched between FM manganites turns into a ferromagnetic metal at 
low temperatures. The induced magnetization in the AFI layer decreases beyond a critical 
width and away from the interface the AFI layer gradually returns to the bulk AFI state. 
This results in an MIT in the FM/AFI superlattice driven by the AFI layer width. Only weak 
external magnetic field is required to tune the insulating into a metallic 
state. The sign of the magnetoresistance changes at higher temperatures. A 
remarkable result is that colossal magnetoresistance in manganite superlattices 
is achieved at lower magnetic fields as in the bulk phases of the constituent 
materials. For the range of AFI layer widths, at which FM/AFI superlattices show a 
metallic behavior at low temperatures, a hump in the temperature dependence of 
the resistivity $\rho$ is observed. The hump in $\rho$ is due to the presence of 
two ferromagnetic transition temperatures in the superlattice. The height of the 
hump decreases and shifts to higher temperatures with decreasing AFI layer width. 

In summary, the width of the AFI layers controls the magnitude of the magnetoresistance 
and the MIT in the FM/AFI superlattices. Our 2D model Hamiltonian calculations provide a 
basis for explaining the MIT in manganite superlattices. The non-monotonic behaviour 
of the induced ferromagnetic moment in the AFI layer is traced to the effects of 
magnetic disorder at the interface. 

\section*{ACKNOWLEDGEMENT}

This work was supported by the Deutsche Forschungsgemeinschaft through TRR80. 
We acknowledge helpful discussions with Sanjeev Kumar.



\end{document}